\documentclass[12pt]{article}

\title{On a "Robust" A-like State of $^3He$ in Aerogel}

\author{G.A. Baramidze and G.A. Kharadze\\
\\
{\it Andronikashvili Institute of Physics, Tamarashvili str. 6,}\\
{\it 0177 Tbilisi, Georgia (e-mail: gogi@iphac.ge)}}

\begin{document}
\maketitle

\begin{abstract}

 The orbitally isotropic Equal Spin Pairing (ESP) state has
been proposed in Ref. [1] as a candidate of an A-like phase of
superfluid $^3He$ in aerogel environment. In order to preserve an
exact isotropy of the state in the presence of the magnetic field
the condensate with equal values of the amplitudes
$\Delta_{\uparrow\uparrow}$ and $\Delta_{\downarrow\downarrow}$
was adopted. Experimentally it is established that this version
does not reproduce observed splitting asymmetry of ESP phase in
aerogel under the action of an external magnetic field. Here we
explore the behavior of the quasi-isotropic version of an
axiplanar ESP phase with $\Delta_{\uparrow\uparrow}\neq
\Delta_{\downarrow\downarrow}$ and show that for this state the
splitting asymmetry ratio could be reconciled with experimental
observations.

\end{abstract}

\vspace{1cm}

In the past decade a problem of the structure of ordered
(superfluid) states of liquid $^3He$ placed in a disordered medium
(aerogel) has attached much attention.

Recently it was pointed out in [1] that the $A$-like phase in
aerogel, which undoubtedly belongs to an ESP category of
spin-triplet condensates, should be different from an axially
anisotropic ABM state because the presence of a spatial disorder
lifts the degeneracy of this phase with respect to the orientation
of orbital anisotropy axes $\hat{l}$, thus preventing the
establishment of a true long-range order (see, also Ref.[2]). In
Ref.[1] it has been proposed to start from ESP order parameter

\begin{equation}
A_{\mu i}=\frac{\Delta}{\sqrt{3}}(\hat{d}_{\mu}a_i+
\hat{e}_{\mu}b_i),~~~~~\hat{d}\bot\hat{e}
\end{equation}
with orbital vectors $\vec{a}=\vec{m}+i\vec{n}$ and
$\vec{b}=\vec{l}+i\vec{p}$, and  to choose four real vectors
$(\vec{m},\vec{n},\vec{l},\vec{p})$ in a way as to satisfy an
orbital isotropy condition

\begin{equation}
Re(A_{\mu i}A^*_{\mu j})=const\cdot\delta_{ij},
\end{equation}
which guarantees that the superfluid state (1) will be "robust"
with respect to local spatial irregularities in aerogel. This is
because the phenomenologically introduced interaction of Cooper
condensate with spatial disorder imposed by aerogel

\begin{equation}
F_{\eta}=\eta_{ij}Re(A_{\mu i}A^*_{\mu j})=0.
\end{equation}
Here a traceless tensor $\eta_{(ij)}\vec({r})$ characterizes the
local action of aerogel environment on the superfluid condensate.
As long as the condition (2) is satisfied, the orbital degeneracy
is preserved and a long-range order of the type described by (1)
can develop. This is in contrast to an orbitally anisotropic ABM
state with an order parameter

\begin{equation}
A_{\mu i}=\frac{\Delta}{\sqrt{2}}\hat{d}_{\mu}
(\hat{m}+i\hat{n}),~~~~\hat{m}\times\hat{n}=\hat{l}
\end{equation}
for which

\begin{equation}
F_{\eta}=-\frac{1}{2}\Delta^2\eta_{ij}\hat{l}_i\hat{l}_j
\end{equation}
and the rotational degeneracy of an orbital anisotropy axis
$\hat{l}$ is lifted locally.

The possibility of realization of an axiplanar orbitally isotropic
ESP phase of type (1) as an equilibrium state is still under debate
[3,4,5]. For the moment the most direct verification of the possible
realization of an orbitally isotropic ESP phase is supplied by the
observation of the splitting of $A$-like phase into $A_1$-like and
$A_2$-like phases under the action of magnetic field. Most generally
an ESP state in a magnetic field is described by an order parameter

\begin{equation}
A_{\mu i}=\frac{1}{2\sqrt{3}}\biggl[\Delta_{\uparrow\uparrow}
(\hat{d}_{\mu}+i\hat{e}_{\mu})(a_i-ib_i)+\Delta_{\downarrow\downarrow}
(\hat{d}_{\mu}-i\hat{e}_{\mu})(a_i+ib_i)\biggl]
\end{equation}
where $\Delta_{\uparrow\uparrow}$ and
$\Delta_{\downarrow\downarrow}$ stand for the amplitudes of Cooper
condensates with $\uparrow\uparrow$ and $\downarrow\downarrow$
spin configurations, respectively. In zero magnetic field
$\Delta_{\uparrow\uparrow}=\Delta_{\downarrow\downarrow}=\Delta$
and (6) reduces to (1). In Ref.[1] even in the presence of
magnetic field it was assumed that
$\Delta_{\uparrow\uparrow}=\Delta_{\downarrow\downarrow}$, and the
adjustment of an ordered state (see Eq.(1)) to an external
magnetic field is realized due to the presence of a spontaneous
magnetic moment $\vec{M}$ of the condensate proportional to an
orbital variable $\Lambda=(\vec{n}\vec{l}-\vec{m}\vec{p})$ and
pointing along the spin quantization axis
$\hat{s}=\hat{d}\times\hat{e}$. According to Ref.[1] the free
energy density of model (1) is given by ($N(0)$ stands for the
quasiparticle density of states at the Fermi level)

\begin{equation}
\frac{F(H)_{Fomin}}{N(0)}=
\biggl(\tau-\zeta\frac{H\Lambda}{3}\biggl)\Delta^2-
\frac{2\Delta^4}{9}\beta_{15}\Lambda^2 + \frac{\Delta^4}{18}
(\beta_{13}+9\beta_{2}+5\beta_{45}),
\end{equation}
where $\tau=(T-T_{co})/T_{co}$ ($T_{co}$ being the critical
temperature in zero magnetic field), linear-in-field term
(proportional to a tiny particle-hole asymmetry coefficient
$\zeta$) stems for the Zeeman energy $\vec{H}\vec{M}$, and
$\beta_{ij}=\beta_{i}+\beta_{j}+\cdot\cdot\cdot$ is a combination
of phenomenological coefficients regulating the fourth order
contribution in $\Delta$.

Minimization of (7) (at $\beta_{15}<0$) shows that the $A_1$-like
ferromagnetic phase (with $\Lambda=1/2$) is realized as an
equilibrium state in a temperature interval

\begin{equation}
\biggl(1+\frac{B}{\beta_{12}}\biggl)\tau_1=\tau_2\leq\tau\leq\tau_2=\zeta
H/6
\end{equation}
with $B=9\beta_2+\beta_3+5\beta_4+4\beta_5$. The $A_1 - A_2$
splitting asymmetry ratio

\begin{equation}
r=\frac{T_{c1}-T_{co}}{T_{co}-T_{c2}}=-\frac{\tau_1}{\tau_2}=
\frac{-\beta_{15}}{\beta_{15}+B},
\end{equation}
which in weak-coupling limit is equal to 0.16. This estimate shows
that $r_{Fomin}$ is too small to be reconciled with recent
experimental observations [6]. This fact poses a question about
the origin of mentioned discrepancy. One way of resolving this
problem is to lift the assumption
$\Delta_{\uparrow\uparrow}=\Delta_{\downarrow\downarrow}$, adopted
in Ref.[1]. In a model with
$\Delta_{\uparrow\uparrow}\neq\Delta_{\downarrow\downarrow}$ an
exact orbital isotropy of an ESP state with an order parameter (6)
is lost (see below), although the orbital anisotropy which appears
is minute (proportional to $\zeta$). Even in the presence of a
tiny orbital anisotropy of an order parameter the spatial
irregularities imposed by aerogel environment will tend the
superfluid state to break up into domains of a finite size with
linear dimension $L$, but in this case $L$ should be large enough
to maintain the coherency of the condensate in the main body of
the system.

In order to realize the above-mentioned approach we start from an
order parameter

\begin{equation}
A_{\mu i}=\frac{1}{2\sqrt{3}}\biggl\{\Delta_{\uparrow\uparrow}(
\hat{d}+i\hat{e})_{\mu}[\hat{m}+i(\hat{n}-\hat{l})]_i+
\Delta_{\downarrow\downarrow}(
\hat{d}-i\hat{e})_{\mu}[\hat{m}+i(\hat{n}+\hat{l})]_i\biggl\},
\end{equation}
where $(\hat{m}, \hat{n}, \hat{l})$ is a triad of mutually
orthogonal unit orbital vectors. The order parameter (10) is a
simple version of a general ESP state. It can be readily shown
that for (10)

\begin{equation}
Re (A_{\mu i}A_{\mu
j}^*)=\frac{1}{6}[(\Delta_{\uparrow\uparrow}^2+
\Delta_{\downarrow\downarrow}^2)\delta_{ij}-
(\Delta_{\uparrow\uparrow}^2-\Delta_{\downarrow\downarrow}^2)
(\hat{n}_i\hat{l}_j+\hat{l}_i\hat{n}_j)],
\end{equation}
which reveals an orbital anisotropy of (10) proportional to
$(\Delta_{\uparrow\uparrow}^2-\Delta_{\downarrow\downarrow}^2)$.
In a standard way a free energy density corresponding to (10) can
be constructed:

\begin{equation}
\frac{F(H)}{N(0)}=\frac{1}{2} (\tau-\zeta H)
\Delta^2_{\uparrow\uparrow}+\frac{1}{2} (\tau+\zeta H)
\Delta^2_{\downarrow\downarrow}+
\frac{1}{4}\beta(\Delta^4_{\uparrow\uparrow}+\Delta^4_{\downarrow\downarrow})+
\frac{1}{2}
\beta^1\Delta^2_{\uparrow\uparrow}\Delta^2_{\downarrow\downarrow},
\end{equation}
where

\begin{eqnarray}
\beta=\beta_{24}+\frac{1}{9}\beta_3,\nonumber\\
\\
\beta^1=\frac{1}{9}(2\beta_1+9\beta_2+\beta_{34}+10\beta_5).\nonumber
\end{eqnarray}

The free energy density (12) coincides in form with that of an
$A$-phase (axial ABM state) for which $\beta=\beta_{24}$ and
$\beta^1=\beta_{24}+2\beta_5$. On the other hand, there is a
crucial difference between ABM state and the state described by
Eq.(10): the former is characterized by strong interaction with
spatial disorder (see Eq.(5)) whereas the latter has only a weak
orbital contact with aerogel structure$\bigl(F_{\eta}=-\frac{1}{6}
(\Delta_{\uparrow\uparrow}^2-\Delta_{\downarrow\downarrow}^2)\eta_{ij}
(\hat{n}_i\hat{l}_j+\hat{l}_i\hat{n}_j)\bigl)$.

In the temperature interval
\begin{equation}
-\frac{\beta+\beta^1}{\beta-\beta^1}\tau_1\leq\tau\leq\tau_1=\zeta
H
\end{equation}
an $A_1$-like phase is realized within a domain of linear size
$L$, and at $\tau\leq\tau_2$ the $A_2$-like phase is stabilized.
For $\Delta^2_{\uparrow\uparrow}$ and
$\Delta^2_{\downarrow\downarrow}$ the following solution is
obtained:
\begin{eqnarray}
\Delta^2_{\uparrow\uparrow}=-N(0)\frac{\tau-\tau_1}{\beta},\nonumber\\
&~~~~~~~~~~~~\tau_2 & \leq \tau \leq\tau_1\\
\Delta^2_{\downarrow\downarrow}=0;\nonumber
\end{eqnarray}
\vspace{0.8cm}
\begin{eqnarray}
\Delta^2_{\uparrow\uparrow}=-N(0)\frac{\tau+\tau_2}{\beta+\beta^1},\nonumber\\
&~~~~~~~~~~~~~~\tau &\leq\tau_2\\
\Delta^2_{\downarrow\downarrow}=-N(0)\frac{\tau-\tau_1}{\beta+\beta^1};\nonumber
\end{eqnarray}
On lowering the temperature below $T_{c1}$ the difference
$(\Delta_{\uparrow\uparrow}^2-\Delta_{\downarrow\downarrow}^2)$
increases gradually and saturates (on reaching $T_{c2}$) at the
level
$(\Delta_{\uparrow\uparrow}^2-\Delta_{\downarrow\downarrow}^2)_{max}=2N(0)\frac{\zeta
H}{\beta-\beta^1}$, so that

\begin{equation}
F_{\eta}^{max}=-\frac{1}{3}N(0)\frac{\zeta
H}{\beta-\beta^1}\eta_{ij}(\hat{n}_i\hat{l}_j+\hat{l}_i\hat{n}_j).
\end{equation}
As mentioned above, the interaction strength of a condensate
having the order parameter (10) with spatial irregularities
imposed by aerogel is very weak due to the presence of a tiny
coefficient $\zeta$. At the same time the $A_1 - A_2$ splitting
asymmetry ration for this case reads as

\begin{equation}
r=-\frac{\tau_1}{\tau_2}=1-\frac{2\beta^1}{\beta+\beta^1},
\end{equation}
and in the weak-coupling approximation (where $\beta^1=0$) the
splitting asymmetry is absent ($r=1$). This fact seems not to be
in conflict with experimental data. \vspace{0.5cm}

We thank G.E. Volovik for correspondence.

\end{document}